\documentclass[epjST]{svjour}
\usepackage{graphicx}
\usepackage{graphics}
\usepackage{psboxit,epsfig,wrapfig,boxedminipage,enumerate} 
\usepackage[varg]{txfonts} 
\usepackage[latin1]{inputenc}
\usepackage{rotating, color}
\usepackage{bm}   

\newcommand{\kt} {k_{\rm B}T}

\newcommand{\bea}{\begin{eqnarray}}
\newcommand{\eea}{\end{eqnarray}}
\newcommand{\vect}[1]{\mathbf{#1}}
\newcommand{\imag}{{\rm i}}

\newcommand{\tr}{\textcolor{black}}

\begin{document}
\title{ Solid phase properties and crystallization in simple model systems  }
\author{F. Turci\inst{1} \and T. Schilling \inst{1} \and M. H. Yamani \inst{2} \and M. Oettel \inst{2}\fnmsep\thanks{\email{martin.oettel@uni-tuebingen.de}} }
\institute{ Universit\'e du Luxembourg, Theory of Soft Condensed Matter, L-1511 Luxembourg, Luxembourg 
  \and Institut f\"ur Angewandte Physik, Universit\"at T\"ubingen, Auf der
  Morgenstelle 10, 72076 T\"ubingen, Germany  }
\abstract{
We review theoretical and simulational approaches
to the description of equilibrium bulk crystal and interface properties as well as to the nonequilibrium processes
of homogeneous and heterogeneous crystal nucleation for the simple model 
systems of hard spheres and Lennard--Jones particles. For the equilibrium properties of bulk and interfaces, density functional theories
employing fundamental measure functionals prove to be a precise and versatile tool, as exemplified with a closer analysis of the
hard sphere crystal--liquid interface. A detailed understanding of the dynamic process of nucleation in these model 
systems nevertheless still relies on simulational approaches. We review bulk nucleation and nucleation at structured walls and examine
in closer detail the influence of walls with variable strength on nucleation in the Lennard--Jones fluid. We find that a planar crystalline substrate induces the growth of a crystalline film for a large range of lattice spacings and interaction potentials. 
Only a strongly incommensurate substrate and a very weakly attractive substrate potential lead to crystal growth with a non--zero contact angle.  
} 
\maketitle
\section{Introduction}
\label{intro}

It is a core interest of statistical mechanics to understand thermodynamic properties of the solid phase
(such as liquid/solid coexistence densities, the equation of state, solid--liquid interfacial tensions...)
from a basic, possibly simple Hamiltonian of the system. Furthermore, homogeneous (in the oversaturated bulk) 
and heterogeneous (at walls, say) nucleation of the solid phase can be studied by microscopic approaches.
Molecular simulation,  density functional theory (DFT) and phase field / phase field
crystal (PFC) models are the main computational approaches to these questions. 
It is hoped that by a thorough understanding of the solid phase thermodynamics and its growth dynamics
in simple model systems such as hard spheres and the Lennard--Jones fluid one gains a sufficient basic knowledge
to understand these issues also for real materials (e.g. metals). Since the materials science community 
often works with the coarse--grained phase field and PFC models, it is desirable to link the microscopic
descriptions of molecular simulation and DFT to the parameters employed in phase field/PFC calculations.

In this paper, we review briefly the density functional descriptions of the equilibrium solid phase
and the solid--liquid interface in simple model systems (Section 2). In Section 3 we focus particularly
on the hard sphere solid--liquid interface where we have obtained a rather complete and consensual  picture through
simulation and DFT, with differences to the PFC description remaining. New DFT results on the metastable
interface of a hard sphere bcc solid/liquid interface are presented. 
Further, we focus on homogeneous and heterogeneous nucleation of the solid phase
in the hard sphere and Lennard--Jones fluids. In these simple model systems, our main source of knowledge
are molecular simulations. In Section 4 we discuss homogeneous nucleation with emphasis on results from the past years.
Section 5 treats heterogeneous nucleation, in particular the crystal growth at planar, crystalline walls.
Through different wall potentials and different wall crystal structure/lattice constant the possible crystallization pathways
are changed both qualitatively and quantitatively. We present new results on Lennard--Jones system at variable
Lennard--Jones type walls. Section 6 concludes our work with a short summary. 


\section{Density functional theory}

Within density functional theory, inhomogeneous liquids and crystals are treated on equal footing,
i.e. the bulk crystal is viewed as a self--sustained oscillation of the one--body density.
In equilibrium, the theory rests on a minimization principle for the grand canonical free energy which is a functional
of this one-body density $\rho(\mathbf{r})$,
\bea
 \Omega[\rho] = \mathcal{F}^{\rm id}[\rho] + \mathcal{F}^{\rm ex}[\rho]-\int d^3{r} (\mu-V^{\rm ext}(\mathbf{r}))\;,
\eea
where $\mathcal{F}^{\rm id}$ and $\mathcal{F}^{\rm ex}$ denote the ideal and excess free energy
functionals of the fluid. $\mu$ denotes the chemical potential 
and the external potential is represented by $V^{\rm ext}$. The exact form of the ideal part
of the free energy is given by
\bea
\label{eq:fid}
\beta \mathcal{F}^{\rm id}[\rho]=\int d^{3}r \beta f^{\rm id}(\mathbf{r})=
\int d^{3}r\rho(\mathbf{r}) ( \ln[\Lambda^{3} \rho(\mathbf{r})]-1 ) \;.
\eea
Here, $\Lambda$ is the thermal de-Broglie wavelength and $\beta = 1/(\kt)$.
The equilibrium density profile  $\rho_{\rm eq}(\mathbf{r})$
is determined via minimizing the grand canonical free energy functional:
\bea
\label{eq:ele}
\beta^{-1}\ln{\frac{\rho_{\rm eq}(\mathbf{r})}{\rho_{0}}} = -\frac{\delta \mathcal{F}^{\rm ex}[\rho(\mathbf{r})]}
{\delta \rho(\mathbf{r})}+\mu^{\rm ex}-V^{\rm ext}(\mathbf{r}).
\eea
For the equilibrium bulk crystal, $V^{\rm ext}(\mathbf{r}) =0$ and $\rho_{\rm eq}(\mathbf{r})$ is lattice--periodic,
and $\rho_{0}$, the homogeneous density (bulk density), is fixed by the excess chemical potential
$\mu^{\rm ex}$.

The central difficulty consists in determining the excess free energy
functional $\mathcal{F}^{\rm ex}$. Only for hard bodies there exists a geometric approach (Fundamental Measure Theory, FMT)
which leads to very precise functionals, for reviews see Refs.~\cite{Tar08,Rot10}. For hard spheres, properties of crystals and crystal--liquid interfaces 
have been examined in sufficient detail such that one may say that we possess a close--to--exact reference theory
for the hard sphere solid. See App.~\ref{app:fmt}  for the explicit form of $\mathcal{F}^{\rm ex}$ for FMT.   

For the one--component, hard sphere (diameter $\sigma$) bulk solid, FMT is in very good agreement with simulations regarding coexistence densities, free energies of the solid and liquid phase,
and density distributions around fcc lattice sites (in particular with regard to their width and anisotropy) \cite{Oet10}.
The equilibrium vacancy concentration found in FMT is smaller than in simulations (2$\cdot$10$^{-5}$ vs. 2$\cdot$10$^{-4}$). 
However, other DFT models and PFC do not predict at all such small concentrations. Furthermore, the correct
description of vacancies restricts the choice of possible fundamental FMT functionals to the White Bear II (tensor) functional
(see App.~\ref{app:fmt}). 

Within FMT, also an accurate description of metastable crystal phases (bcc, hcp) is possible \cite{Lut06a,Yam13}. For densities
$\rho_0\sigma^3= \rho_0^*> 1.16$, two metastable bcc phases are found. Their relevance remains to be investigated possibly in applications
to solids with stable bcc phases (with FMT as a reference theory). The hcp phase is actually more stable than fcc in FMT (by a tiny free
energy difference of 0.001 $\kt$ per particle). In simulations, the situation is precisely reverse (also by about the same free energy difference),
this can be shown to be a consequence of subtle multi--body correlation effects which are missed by FMT \cite{Yam13}.    

The phase diagram of binary hard spheres has been investigated in Ref.~\cite{Son08}, again finding good agreement with simulations. 

Previous studies of the crystal--liquid interface in FMT involved restricted parametrizations of the three--dimensional density profile across the
interface \cite{Lut06b,Son06}. This introduces some uncertainty as to the precision for the value of the interfacial tension since the functional
minimization is constrained. Depending on the particular FMT functional the interfacial tension is close \cite{Lut06b} or 25\% above \cite{Son06} corresponding
simulation values. In Section 3 we discuss the hard sphere crystal--liquid interface in more detail and present also results of unconstrained 
minimizations.   

\subsection{Functional Taylor expansions}
\label{sec:taylor}

Many practical applications of DFT
have started from an expansion of ${\cal F}^{\rm ex}$ around a background reference density profile $\rho_0(\vect r)$ which,
in general, can depend on the position $\vect r$:
\bea
 \label{eq:f_hnc}
 \beta {\cal F}^{\rm ex} = \beta F^{\rm ex}_0[\rho_0] - \int d^3 r c^{(1)}(\vect r;\rho_0)\Delta\rho(\vect r) - \frac{1}{2} \int d^3 r d^3 r'
    c^{(2)}(\vect r, \vect r';\rho_0) \Delta\rho(\vect r)\Delta\rho(\vect r') + \dots
\eea
Here, $F^{\rm ex}_0[\rho_0]$ is the excess free energy pertaining to the background profile,
$\Delta\rho(\vect r) = \rho(\vect r)-\rho_0(\vect r)$ and $c^{(1)}$ and $c^{(2)}$ are the first two members in the hierarchy
of direct correlation functions $c^{(n)}$, defined by
\bea
 \label{eq:cn_def}
 c^{(n)}(\vect r_1,\dots,\vect r_n;\rho_0) = -\beta \left.
   \frac{\delta^{(n)} {\cal F}^{\rm ex}}{\delta \rho(\vect r_1)\dots \delta \rho(\vect r_n)}
   \right|_{\rho=\rho_0(\vect r)} \;.
\eea
In most practical applications, $\rho_0\equiv const.$ is taken to be a reference bulk density in which case
$-c^{(1)} = \beta \mu^{\rm ex}= \beta\mu-\log(\rho_0\Lambda^3)$ and $c^{(2)}(\vect r-\vect r';\rho_0)$ depends only on
{the coordinate difference of the two positions $\vect r$ and $\vect r'$}. To evaluate the functional in Eq.~(\ref{eq:f_hnc}), 
the correlation function $c^{(2)}$ has to be determined as an external input, 
provided e.g.\ by integral equation theory or by simple approximations of RPA type \cite{Eva79}.

The work of Ramakrishnan and Yussouff on the Taylor--expanded functional applied to hard spheres initiated the density functional research on freezing
\cite{Ram79}. However, only for soft systems (Gaussian particles as a model for e.g. polymers or dendrimers) the functional is reliable
\cite{Pin12}. If applied to other soft systems, the Taylor--expanded functional works better for repulsive systems, see e.g.
Ref.~\cite{Hei13} for a recent study on the Yukawa model. In the studies mentioned, the employed direct correlation function $c^{(2)}(\vect r- \vect r';\rho_0)$
has been always the isotropic one from the bulk fluid. Significant improvement of quantitative accuracy can be achieved if one allows
for a direct correlation function containing an anisotropic contribution with the proper crystal lattice symmetry \cite{Sin11,Sin13}. 

If the system under study can be approximated by a reference system with corresponding functional
${\cal F}^{\rm ex, ref}$, then the expansion (\ref{eq:f_hnc}) can be modified to
\bea
 \label{eq:f_refhnc}
 \beta {\cal F}^{\rm ex} &=& \beta {\cal F}^{\rm ex, ref} + \beta \Delta F^{\rm ex}_0[\rho_0] - \int d^3 r \Delta c^{(1)}(\vect r;\rho_0)\Delta\rho(\vect r) - \\
 & &  \frac{1}{2} \int d^3 r d^3 r'
    \Delta c^{(2)}(\vect r, \vect r';\rho_0) \Delta\rho(\vect r)\Delta\rho(\vect r') + \dots \nonumber
\eea
where the expansion coefficients reflect differences between the actual and reference system:
$\Delta F^{\rm ex}_0[\rho_0] = F^{\rm ex}_0[\rho_0]- F^{\rm ex,ref}_0[\rho_0]$ for the free energy difference pertaining to 
the profile  $\rho_0$ and in a similar fashion $\Delta c^{(1)}$ and $ \Delta c^{(2)}$ are defined.
Since from a practical point of view only hard spheres qualify for a suitable reference system with a sufficiently precise reference 
functional, this approach works for simple fluids with repulsive cores such as Lennard--Jones or Yukawa fluids, or the primitive
model for electrolytes \cite{Gil03}.

The first quantitative description of the Lennard--Jones phase diagram in this spirit was given in Ref.~\cite{Cur86}, still with
the older weighted--density approach for the hard sphere reference functional. Only recently, a study employed FMT as a reference functional and found 
remarkably precise values for the crystal--liquid binodal and crystal--liquid interfacial tensions of the Lennard--Jones system \cite{Wan13}.

\subsection{Phase field crystal (PFC) model}
\label{sec:PFC}

The Taylor expanded functional in Eq.~(\ref{eq:f_hnc}) is nonlocal in the densities. Through an additional approximation (gradient
expansion) it can be cast into a local form. We consider a constant reference density $\rho_0$ and the following
power expansion of the Fourier transform of the direct correlation function:
\bea
   \tilde c^{(2)} (k;\rho) = -c_0 +c_2\; k^2 - c_4\;k^4 \dots
\eea
Using this, the Taylor--expanded functional becomes
\bea
   \beta {\cal F}^{\rm ex}_{\rm loc} = \beta F^{\rm ex}_0(\rho_0) + \beta \mu^{\rm ex} \int d^3 r \Delta \rho(\vect r) +
             \frac{1}{2} \int d^3 r \Delta\rho(\vect r)\left( c_0 + c_2\nabla^2 + c_4 \nabla^4 \dots   \right)\Delta\rho(\vect r)
    + \dots \nonumber \\
\eea
We observe that the excess free energy density contains local terms up to order 2 in $\Delta\rho$ and up to order 4 in $\nabla(\Delta\rho)$.
The total free energy contains in addition the ideal gas term, ${\cal F}^{\rm id}[\rho]$ from Eq.~(\ref{eq:fid}).
One may expand also this term in $\Delta\rho$ in order to obtain a power--expanded free energy density
to power 4 in both $\Delta\rho$ and $\nabla(\Delta\rho)$.
It turns out that the phase diagram of such a power--expanded model  is equivalent to 
a reduced model with a dimensionless free energy according to \cite{Jaa10}
\bea
 \label{eq:fpfc}
  F_{\rm PFC} = \int d^3 x f_{\rm PFC} = \int d^3 x \frac{1}{2} \left( \Psi(\vect x)
                 \left[ -\epsilon + (1+\nabla^2)^2 \right] \Psi(\vect x) +  \frac{\Psi(\vect x)^4}{4} \right),
\eea
which we call the phase field crystal (PFC) model. Here, $\vect x  =  q_0 \vect r$ is a dimensionless coordinate
with $q_0= \sqrt{\frac{c_2}{2c_4}}$ being the wavenumber of ``favored'' density oscillations.
$\Psi=\rho/\rho_0-b$ is a reduced and shifted density. The parameter $\epsilon$
(playing the role of a temperature), the shift $b$ and the free energy scale of the model can be related 
to the power expansion parameters of the Taylor--expanded free energy. 
However, the phase diagrams of a simple fluid (like hard spheres or Lennard--Jones) and PFC cannot be mapped
onto each other by this route. Rather, the power expansion parameters of the Taylor--expanded free energy
should be treated as fitting parameters and then linked to the PFC model phase diagram.

The PFC model is a very generic model with possible periodically ordered equilibrium states such as 
stripes, rods and bcc, fcc, hcp \cite{Low12}. This generic nature offers many possibilities to understand structure formation processes
qualitatively, care has to be taken when a quantitative understanding is desired. For the case of iron (bcc), bulk properties,
interfacial tensions and anisotropies are obtained in reasonable agreement with simulations \cite{Wu07,Jaa09}. For the case of a Yukawa fluid (bcc),
interfacial tension are too large by a factor of 2 and anisotropies are greatly exaggerated \cite{Hei13}.
An attempt to relate the PFC pair correlations in an amorphous solid (bcc part of the PFC phase diagram) to a pair potential
yielded a strongly oscillatory potential with a rather soft core \cite{Gra11}. This potential is very much unlike the Yukawa potential
or a possible potential for iron. 
 For the case of hard spheres
(fcc), the fit of bulk properties resulted in interfacial tensions too low by more than a factor of 3 and interface profiles for $\Psi(\vect r)$
which could not be related to the very precise FMT density profiles \cite{Oet12b}.

\section{The hard--sphere crystal--fluid interface}

The density profile and the interfacial tension $\gamma$ of the hard--sphere crystal(fcc)--fluid interface have been under intense 
scrutiny by DFT and simulation approaches. Within the older weighted--density approach, Ref.~\cite{Ohn94} demonstrated 
that a full minimization with respect to the three--dimensional (3d) density profile leads to values significantly lower ($\beta \gamma \sigma^2 =\gamma^* \sim 0.3$)
than obtained by restricted minimizations of parametrized profiles (e.g. $\gamma^*  \sim 0.6$ in Ref.~\cite{Cur89}). 
This underlines the need to perform also full minimization in the case of FMT functionals, besides the restricted
minimization of Refs.~\cite{Lut06b,Son06}. In Ref.~\cite{Har12} such results for the WBII--T functional (see App.~\ref{app:fmt})
are reported with the following results for the orientation--dependent interfacial tension:
$\gamma^*_{[100]}=0.69$, $\gamma^*_{[110]}=0.67$ and $\gamma^*_{[111]}=0.64$. The molecular dynamics results reported in the same
Ref.~\cite{Har12} are $\gamma^*_{[100],\mbox{sim}}=0.64$, $\gamma^*_{[110],\mbox{sim}}=0.62$ and $\gamma^*_{[111],\mbox{sim}}=0.60$.  
Whereas the tension anisotropies from DFT and from the simulation agree, the overall values from the simulations are somewhat smaller,
presumably due to capillary wave effects (see below). We remark that the extraction of crystal--liquid interfacial tensions
from simulations is still a delicate case, for hard--spheres reported values differ by 10\% \cite{Dav10,Fer12,Har12}. 

Useful insights into the structure of the interface may be obtained by considering a mode expansion \cite{Oet12a}.
Let the density field be $\rho(x,y,z)$ which describes the crystal--fluid interface with interface normal
in $z$--direction.  We can parametrize it
in terms of a modified Fourier expansion
\bea
 \rho(x,y,z) = \sum_j \exp(\imag \vect K_j \cdot \vect r)\; p_j(z)\;,
\eea
where $\vect K_j$ denotes the reciprocal lattice vector (RLV), $j$ and
the $z$--dependent Fourier amplitude $p_j(z)$ are modes of the field.
One expects that upon
crossing the interface  from the crystal side, all $p_j(z)$ relax to zero for nonzero $\vect K_j$. Only
for $\vect K_j \equiv 0$, the value for the associated mode $p_0$ crosses from the average crystal density $\rho_{\rm cr}$ of the crystal to
the average fluid density  $\rho_{\rm fl}$ at coexistence.

Properties of the interface modes as obtained from FMT have been discussed in detail in Ref.~\cite{Oet12a}.
The most important conclusions are:
\begin{enumerate}
 \item approaching the interface from the liquid side, crystallinity sets in earlier as densification: there is
       a separation of about one cubic unit cell length $a \approx 1.6$ $\sigma$
       between  the interface location as determined by the average density
       and the interface location as determined by the leading crystallinity mode ($p_{1}(z)$ with $\vect K_1 =(2\pi/a)(1,1,1)$)
 \item a small density depletion zone just in front of the bulk crystal (dip in profile $p_{0}(z)$)
 \item strongly non-monotonic mode profiles also for next--to--leading modes, especially for
       $p_{2}(z)$ with $\vect K_2 =(2\pi/a)(0,0,2)$ (leading crystallinity mode for lateral density average)
\end{enumerate}

\begin{figure}
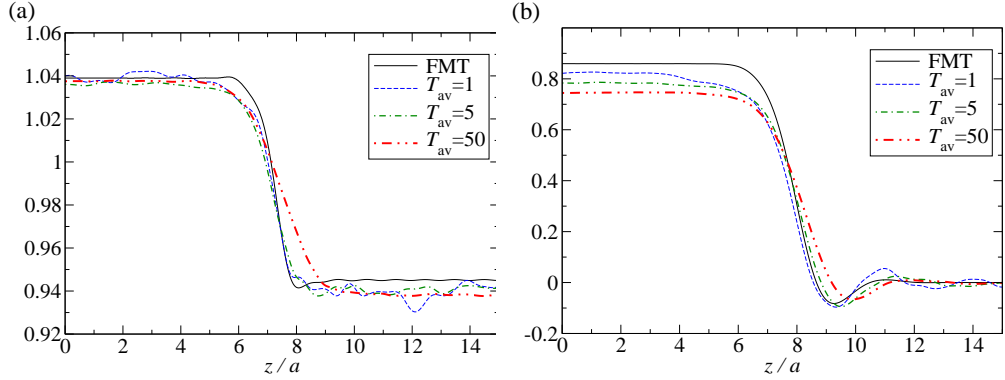

  \epsfig{file=fig1a.eps, width=6.5cm} \hspace{0cm}
  \epsfig{file=fig1b.eps, width=6.5cm} 
  \caption[]{ Leading modes extracted from MD simulated, laterally averaged density profiles in comparison with FMT results.
 (a) average density mode $p_{0}(z)$, (b) real part of $p_{2}(z)$ with $\vect K_{2}=(2\pi/a)(0,0,2)$.
 \tr{The averaging time in the MD simulations was about one self--diffusion time. For longer times, clear effects of capillary wave 
    broadening were seen which are absent in FMT. See also Ref.~\cite{Zyk10}. }
 Figure adapted from Ref.~\cite{Oet12b}.    }
  \label{fig:modes_sim}
\end{figure}

The leading modes have been also extracted from molecular dynamics simulations and found to be in good agreement with FMT 
\cite{Oet12b}. 
As an example, in Fig.~\ref{fig:modes_sim} we present a comparison between FMT 
and simulation for the average density mode $p_0$ and the leading crystallinity mode for lateral density average $p_2$.
Simulation results are presented for different averaging times $T_{\rm av}$, given in units of the characteristic self--diffusion
time (time it takes a particle in
the coexisting liquid to diffuse over a distance of $\sigma$).   
For $T_{\rm av}\sim 1$ we observe nearly quantitative agreement, whereas longer averaging times lead to broadening effects due 
to capillary waves (not captured in FMT).  

\begin{table}
\begin{center}
\begin{tabular}{llllllll}
   \hline \hline 
     & functional   &  $\rho_{\rm fl}^*$ & $\rho_{\rm cr}^*$ &  $n_{\rm vac}$ & $\gamma_{\rm [100]}^*$ & $\gamma_{\rm [110]}^*$ & $\gamma_{\rm [111]}^*$ \\ \hline 
bcc &RF--T     & 0.923 & 0.971 & 0.012 & 0.25 & 0.25 & 0.24 \\
    &WB--T     & 0.976 & 1.012 & 0.013 & 0.27 & 0.26 & 0.25 \\
    &WBII--T     & 1.016 & 1.045 & 0.015 & 0.34 & 0.33 & 0.32 \\ \hline
fcc &WBII--T      & 0.945 & 1.039 & 2$\cdot$10$^{-5}$ & 0.69 & 0.67 & 0.64 \\ \hline \hline
\end{tabular}
\end{center}
\caption{Coexistence properties for bcc in comparison with fcc for different functionals (given explicitly in App.~\ref{app:fmt}). 
Note the rather high equilibrium vacancy concentration $n_{\rm vac}$ of bcc as well as the strongly reduced interfacial tension values
when compared with fcc. Details of the applied numerics are described in Refs.~\cite{Oet12a,Yam13}. }
\label{tab:bcc}
\end{table} 

DFT allows to study of metastable crystals \cite{Lut06b,Yam13} as well as the associated crystal--liquid interfaces.
As a novel result, we present fully minimized FMT results for bcc coexistence properties and bcc--liquid interfacial tensions in Tab.~\ref{tab:bcc},
obtained with the numerical procedure described in Ref.~\cite{Oet12b}.
First, we observe a rather large discrepancy in the values of the coexistence densities when comparing the different functionals.
(The functionals are described in App.~\ref{app:fmt}.)
In case of functional RF--T the lower coexistence densities result from the underlying Percus--Yevick equation of state for the
liquid phase which overestimates the pressure near coexistence. For the liquid phase, functionals WB--T and WBII--T give nearly identical results,
yet noticeable differences for the crystals exist (notice that WB--T also does not predict a nonzero vacancy concentration $n_{\rm vac}$
for the fcc crystal \cite{Oet10}). In view also of the higher degree of consistency in constructing the WBII--T functional, we 
think that it is the most suited functional for crystals. The bcc interfacial tension values are smaller by a factor of 2 and more
compared with the fcc values. Although this means that nucleation of bcc should be easier than fcc,
the difference in $F/N$ around fcc--liquid coexistence
($\rho_{\rm cr}^* \approx 1.04$) is very high with about 0.3 $\kt$ \cite{Yam13},  therefore bcc nucleation should be inhibited.     
The bcc solutions are furthermore a useful reference point for discussing the crossover from fcc to bcc as the most stable crystal
structure
for other potentials such as of $(\sigma/r)^n$ type. Also, a crossover for the bcc interfacial tension from
metastable to stable interfaces can be discussed using the reference functional approach (see Eq.~(\ref{eq:f_refhnc}))
\cite{War10,War12}. By going the reverse way (using simulated bcc interfacial tension values for the $(\sigma/r)^n$--potential), 
Ref.~\cite{War12} estimated the reduced hard sphere bcc interfacial tension to be around 0.4, in reasonable agreement with our WBII--T results.

\tr{
Are stable and metastable HS crystal--liquid interfaces relevant for the study of metals? 
One can define a coefficient $\alpha$ (Turnbull coefficient) through the relation 
$\gamma \rho_{\rm cr}^{-2/3} = \alpha\, \Delta H_f$ where $\Delta H_f= (H_{\rm cr} - H_{\rm fl})/N$ is the enthalpy of fusion (enthalpy difference
between coexisting crystal and liquid per particle). MD simulations for various metals using embedded atom potentials showed that in a 
$\gamma \rho_{\rm cr}^{-2/3}$--$\Delta H_f$ plot fcc metal points are close to a straight line with $\alpha \approx 0.55$ and bcc metal points
are close to a straight line with $\alpha \approx 0.29$ \cite{Hoy04}. For hard spheres, 
$\Delta H_f= (F_{\rm cr} - F_{\rm fl})/N = p_{\rm coex}(1/\rho_{\rm fl}-1/\rho_{\rm cr})$ and our FMT values for fcc 
give $\alpha \approx 0.6$, close to the fcc metal value \cite{Lai01}. Thus just entropy determines the Turnbull coefficient, attractions in the system
change $\gamma \rho_{\rm cr}^{-2/3}$ and $\Delta H_f$ in the same proportion. For bcc, the situation is less clear: between the three functionals we have used,
$\alpha$ differs from 0.40 (RF--T), 0.52 (WB--T) to 0.73 (WBII-T), mainly due to the differences in the coexistence gap $\rho_{\rm cr}-\rho_{\rm fl}$.
Thus, our tentative conclusion would be that attractions in metals (i) stabilize the bcc phase and (ii) also change 
$\gamma \rho_{\rm cr}^{-2/3}$ and $\Delta H_f$ disproportionately to give a lower Turnbull coefficient.        
Note, however, that experimental studies on metastable metal droplets seem to give Turnbull coefficients which are systematically larger than 
the simulated values ($\alpha_{\rm fcc} \approx 0.6 ... 0.8$, $\alpha_{\rm bcc} \approx 0.6$) \cite{Her10}.  
}

\section{Simulation of homogeneous crystal nucleation}
At a first order phase transition, such as the liquid--to--crystal transition, 
the phase transformation process is subject to kinetic barriers. Meta--stable 
phases can persist over long times and then suddenly and quickly transform 
by nucleation and growth into the stable phase \cite{Oxtoby1997}. 
Nucleation at a first order phase transition is traditionally described by 
classical nucleation theory (CNT) or extensions thereof. 
The nucleation rate density $I$ is assumed to have the form 
\begin{equation}
I = \kappa \exp(-\beta\Delta G^*) \quad ,
\end{equation}
where  $\Delta G^*$ is
the height of the free energy barrier associated with the formation 
of a critical nucleus and $\kappa$ is a kinetic prefactor. The basic 
assumption underlying this type of description is a separation in 
time-scales between slowly varying coordinates such as the size or shape 
of the nucleus and the remaining coordinates that are considered 
thermally equilibrated at all times, constituting a ``free energy landscape'' 
in which the process evolves.  In the most basic version of CNT, 
{the nucleus 
is assumed to be spherical (i.e.~for crystalline nuclei, the anisotropy of 
the interfacial tension is neglected), and  }
the radius $R$ of the nucleus is considered to be the only relevant slow 
coordinate
\begin{equation}
\Delta G(R) = \frac{4}{3} \pi R^3 \rho \Delta \mu + 4 \pi R^2 \gamma \quad ,
\end{equation}
where $\rho$ is the number density of the stable phase at equilibrium, 
$\Delta \mu$ is the chemical potential difference between the meta-stable 
phase and the stable phase and $\gamma$ is the interfacial tension of the 
planar interface between the two phases.

Classical nucleation theory can directly be tested by means of computer 
simulation, because interfacial tensions, supersaturations, nucleation 
rates and cluster morphologies can be computed independently from one 
another. There is a large body 
of literature on simulations of homogeneous nucleation in specific materials, 
which to review is beyond the scope of this 
article. Here we discuss crystal nucleation in two simple model 
systems, hard spheres and the Lennard--Jones system:

Homogeneous crystal nucleation in colloidal 
suspensions of monodisperse hard spheres has been studied extensively
in experiments and simulations over the past 20 years \cite{Schaetzel1993,He1996,Harland1997,Auer2001,Iacopini2009,Filion2010,Schilling2010,Schilling2011,Russo2012}. 
Hard spheres interact only by 
excluded volume, the liquid-to-crystal phase transition in this system 
is purely entropic.
Thus one could expect the transition dynamics to be particularly simple. 
Despite the simplicity of the model system, however, the nucleation rates obtained by computer simulation differ significantly from those observed in experiments. In 2001 Auer and Frenkel used umbrella sampling to compute the crystal nucleation rate, i.e.~an approach based on transition state theory \cite{Auer2001}. As a reaction coordinate to guide the sampling they chose the size of the largest crystalline cluster in the system. Filion and coworkers tested this approach by comparison to unbiased molecular dynamics simulations as well as forward-flux sampling 
(a steady state rare event sampling technique) and concluded that the 
nucleation rates coincided within the error bars. Hence the assumption of 
a separation in time-scales holds, transition state theory can be applied, 
and the source of the discrepancy between 
simulation and experiment must lie elsewhere. We simulated hard spheres with 
Newtonian as well as Brownian dynamics \cite{Schilling2010,Schilling2011} 
and found for both cases a pre-crusor mediated process in which first dense 
aggregates formed which then crystallized. (A similar process has also been 
seen in the Lennard-Jones system \cite{Trudu2006}.) This finding would imply 
that an approach based on biasing with respect to a
reaction coordinate that only takes crystallinity into account should not 
produce the same result as a direct simulation. However, the differences 
due to this effect are probably smaller than the accuracy of the rates in
Ref.~\cite{Filion2010}.

A major difference between experiment and simulation on colloidal suspensions is the presence of a solvent in the experiment. We have recently simulated the nucleation process taking solvent hydrodynamics into account by means of Multi Particle Collision Dynamics and found that the nucleation rates depend strongly on the solvent viscosity (beyond the trivial slowing down of diffusion with increasing viscosity)\cite{Marc2013}. Due to hydrodynamic interactions, spheres are attached cooperatively to the nucleus and hence the nucleation rate is enhanced in viscous solvents. This effect might explain the discrepancies in nucleation rates observed so far. {Furthermore, it shows that it is in general difficult to compare phase transition kinetics of colloids and metals, as colloids are affected by solvent kinetics, even though the equilbrium phase behaviour of the two classes of systems might be similar.} 

Homogeneous crystal nucleation from the undercooled Lennard-Jones melt has 
also been addressed in many simulation studies over the past 10 years 
\cite{Trudu2006,Moroni2005,Wang2007,Peng2008,Lundrigan2009,Baidakov2010,Peng2010,Baidakov2012,Prestipino2012} using a variety of simulation methods 
ranging from free energy based approaches such as umbrella sampling over 
non--equilibrium rare event techniques such as transition path sampling to 
``brute force'' molecular dynamics. In summary, no clear picture 
has emerged yet regarding the range of applicability of CNT. The simulations 
are subject to various finite-size effects \cite{Peng2010}, thus despite the 
simplicity of the model system, it is non-trivial to draw comparisons between 
results from different studies. Several sources for the 
deviation of nucleation rates from the CNT predictions have been identified 
and corrections have been in incorporated into the theory to take into 
account the fact that the interface is 
not sharp and that it fluctuates \cite{Prestipino2012}. But a general coarse 
grained description still remains to be derived.

\section{Simulations results on heterogeneous nucleation and growth on strained structured surfaces}
The CNT approach to homogeneous nucleation can be extended in a straightforward way to take into account the presence of impurities, defects or fixed boundaries of the crystallizing system. Substrate surfaces and localized defects often trigger and accelerate the formation of a crystalline nucleus, because they lower the free energy barrier for nucleation. Extending the classical nucleation scenario, the \textit{heterogeneous nucleation} barrier can be written as \cite{Turnbull:1950iq}

\begin{equation}
\Delta G_{\rm het}=\gamma_{\rm cf}A_{\rm cf}+(\gamma_{\rm cs}-\gamma_{\rm fs})A_{\rm cs}-n\Delta\mu+\tau L
\label{HetCNT}
\end{equation}
In this expression $ \gamma_{\rm cf},\gamma_{\rm cs},\gamma_{\rm fs}$ are the crystal--fluid, crystal--substrate, fluid--substrate interfacial tensions respectively. $A_{\rm cf}$ and $A_{\rm cs}$ are the contact surface areas between crystal and fluid, and crystal and substrate. $\Delta\mu$ is the chemical potential difference between the metastable fluid and the crystal and $n$ the number of crystallized particles. $\tau$ and $L$ are the line tension of the contact line between the substrate, the crystal and the fluid, and the length of this line respectively.
The free energy difference depends sensitively on the curvature of the substrate surface exposed to the growing crystals. Thus impurities, curved substrates, flat unstructured and structured substrates induce different kinetics for nucleation and growth.

As in the case of homogeneous nucleation, computer simulation allows to test the assumptions that enter the simplified, coarse-grained description. But even for simple model systems, the parameter space is large, as substrate curvature and structure have to be taken into account. Hence a variety of different scenarios have been simulated, but as far as we are aware there has not yet been a systematic study to map out the range of applicability of the free energy landscape based, coarse-grained approach to heterogeneous crystal nucleation.

The role of seeds and impurities has been investigated in simple model systems under different conditions, such as microgravity \cite{Schope:2011vt}, in the presence of large spherical impurities in colloidal suspensions \cite{Villeneuve:2005il}, and with variable impurity size through Monte-Carlo simulations \cite{Cacciuto:2004kc,Jungblut:2013gs}. A recent transition path sampling analysis \cite{Jungblut:2013gs} has shown that pre-structured minimal crystalline seeds commensurate with the bulk stable crystal phase enhance the crystallization rate by many orders of magnitude while incommensurate ones have no effect.

Heterogeneous crystallization on planar surfaces has been extensively studied due to its simplicity for several model systems \cite{Hoogenboom:2003jpa,Sandomirski:2011tf,2004PhRvL..93j8303D,Volkov:2002ta}. In the case of colloidal suspensions of hard spheres, unstructured flat substrates induce the formation of oriented crystals with close-packed planes parallel to the substrate \cite{Heymann:1998dh} and simulations have demonstrated that hard spheres next to a flat hard wall have to overcome a small free energy barrier for nucleation \cite{2003PhRvL..91a5703A}, consistently with the pre-wetting transition observed for the contact of a stable fluid phase with a flat wall \cite{2005JPCM...17S.429E}. 

Confinement between flat plates can furthermore induce a rich phase behavior, with fluid-solid transitions dominated either by capillary freezing or melting depending on the spacing between the plates \cite{Fortini:2006cm}. Curved surfaces have recently been the object of theoretical \cite{Garcia:2013jg} and  experimental investigation \cite{Ziese:2013fj}, where the additional frustration introduced by the curvature inhibiting crystal formation can be partly reduced using different topological patterning and defects on the substrate surface.  Density functional theory has also been used to reproduce qualitative aspects of heterogeneous crystallization in the vicinity of a variety of flat and curved substrates \cite{Kahl:2009bm}.

Structured templates provide a large variety of possible template-fluid interactions \cite{2005JPCM...17S.429E}:  striped chemically activated walls  with colloidal suspensions \cite{2002PhRvE..65d1602H}, charged or hard fixed ions with ordered or disordered patterns \cite{Hermes:2011ck}, the latter also used in order to test glassy dynamics and binary mixtures correlation lengths \cite{Scheidler:2004tx,2012PhRvE..85a1102B}. 
Different types of patterned substrates for hard spheres have been considered in \cite{Xu:2010kb}, showing that the substrate can induce and eventually stabilize phases that are normally unstable in the bulk system (such as bcc crystal structures for hard spheres). 

On the theory side,
the crystallization process has also been tested by means of the PFC model \cite{Toth:2012th} (where, as described in Sec.~\ref{sec:PFC}, the
connection to an actual system remains unclear). Very recently, an FMT--type density functional has been employed to
study the nucleation of hard disks on a patterned substrate \cite{Neuhaus:2013bf}. The nucleation 
on the surface follows a so--called \textit{compatibility wave} scenario with preferred growth directions that give the
least mismatch between the 2d substrate and the 2d crystal lattices.

\subsection{Flat stretched substrates: the role of attractive forces}
In order to vary the interfacial tensions in Eq.~(\ref{HetCNT}) 
we simulated crystallization on a planar crystalline substrate that was 
stretched resp.~compressed with respect to the coexistence lattice constant.
Deformation of the crystalline substrate leads to different kinetic pathways 
towards crystal growth.
In Ref.~\cite{Dorosz:2012ka} we discussed crystallization in suspensions of hard spheres and found a crossover between a regime of instantaneous film growth at small substrate lattice spacing (high substrate densities) and a nucleation regime with long induction times at large spacings (low substrate density). The transition is conjectured to occur when the substrate packing fraction is close to that of the packing fraction of the bulk spinodal instability.\footnote{\tr{Note that spinodals are a somewhat ill--defined
mean-field concept in the case of short--range forces \cite{Bin84}. In fact, in Ref.~\cite{Dorosz:2012ka} the spinodal density was estimated through (mean--field) FMT to be the density at which no local
free energy minimum belonging to a crystalline density profile could be found anymore.}}  
Recent experiments appear to confirm such a scenario for colloidal hard spheres \cite{Schoepe2013}. 

Here we present simulations of crystallization in a Lennard--Jones system, in 
which we systematically vary the interaction between the fluid and the 
substrate, to test whether the effect observed in hard spheres 
\cite{Dorosz:2012ka} is also present in attractive systems.

\begin{figure}[h!tbp]
\centering
\includegraphics[width=\textwidth]{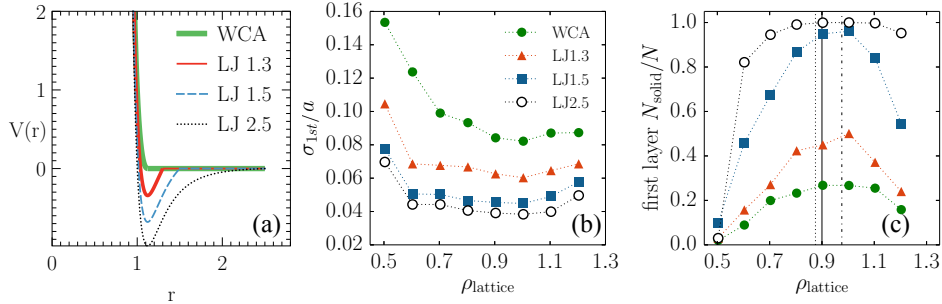}
\caption{(color online) (a) Plot of the wall--fluid interaction potentials considered in this work.  (b) Width of the first density peak close to the substrate in the laterally averaged density profile, $\langle \rho(z)\rangle_{x,y}$,  modeled by a Gaussian function of width $\sigma_{1st}$ at the early stages of growth ($t=0.25 \tau_{D}$): for less attractive forces as well as substrate densities outside of the coexistence region the peaks become broader. (c) Average fraction of solid particles in the first crystalline layer at time ($t=0.25 \tau_{D}$, out of 10 independent growth trajectories. Error bars are within the symbol sizes. The different curves indicate different interactions between the Lennard--Jones metastable fluid at density $\rho=0.95$.The transition between low lattice densities and high lattice densities crosses an optimal value for the solid fraction which is located in the coexistence region. A smooth transition occurs for lattice densities crossing the coexistence region (delimited by the dotted and the dashed--dotted vertical lines), where the spinodal density (continuous vertical line) is located.
}
\label{composition1}
\end{figure}

To do so, we considered a Lennard--Jones fluid in the isochoric, isothermal ensemble confined by some layers of fixed particles forming the patterned substrate (see App. \ref{simulationdetails} for the details of the simulation). The interaction between the substrate and the fluid is tuned changing the cutoff radius of the Lennard--Jones interaction, reducing it from the the standard $2.5\sigma$ radius to the limiting case of the purely repulsive Weeks--Chandler--Andersen (WCA) potential (see Fig.~\ref{composition1}(a)). We track the progression of the crystalline front using the so called average local bond order parameters $\bar{q}_{4},\bar{q}_{6}$ (see App. \ref{simulationdetails}), allowing for the distinction between different crystalline structures such as bcc, hcp and fcc.

In order to distinguish in a quantitative manner between the instantaneous growth and the long induction--time regime associated with heterogeneous nucleation, we monitor the very first steps of the molecular dynamics leading to the formation of a crystalline layer on top of the fixed particles of the substrate. We isolate then the particles confined in the region between the substrate and the first minimum of the density profile and study the fraction of crystalline particles in the first layer. All particles with $\bar{q}_{6}> 0.36$ are regarded as crystalline, consistently with previous LJ calculations \cite{Lechner:2008vz}.

We test a broad range of lattice densities for the substrate, from $0.5$ to $1.2 \sigma^{-3}$, crossing the coexistence region bounded by the coexistence densities $\rho_{\rm fl}=0.8751 \sigma^{-3}$ and $\rho_{\rm cr}=0.9759\sigma^{-3}$ as given by the Lennard--Jones equation of state discussed in \cite{2007JChPh.127j4504M}. For the value of the spinodal transition we refer to a linear interpolation of the data provided in \cite{Kuksin:2007kq}, resulting in $\rho_{\rm spinodal}\approx0.90\sigma^{-3}$ at $k_{B}T=0.8\epsilon$. 
Note that for a fluid with purely repulsive WCA interactions the liquid--solid coexistence region is shifted towards higher densities, lying between 0.92 and $0.99\sigma^{-3}$\cite{Ahmed:2009eg}, thus simulations at equal densities of the supersaturated fluid correspond to slightly different chemical potential differences. Notice moreover that the substrate structures are all but typical commensurable structures for the bulk fluid at the chosen temperature and therefore they represent a template for growth which is in general suboptimal. 

The fluid is initially equilibrated separately, and we use its equilibrium self-diffusion constant $D$ \cite{Meier:2004fp} in order to obtain the self-diffusion time $\tau_{D}=\sigma^{2}/D$ used as the unit of time in the rest of the analysis.

\begin{figure}[h!tbp]
\centering
\includegraphics[width=\columnwidth]{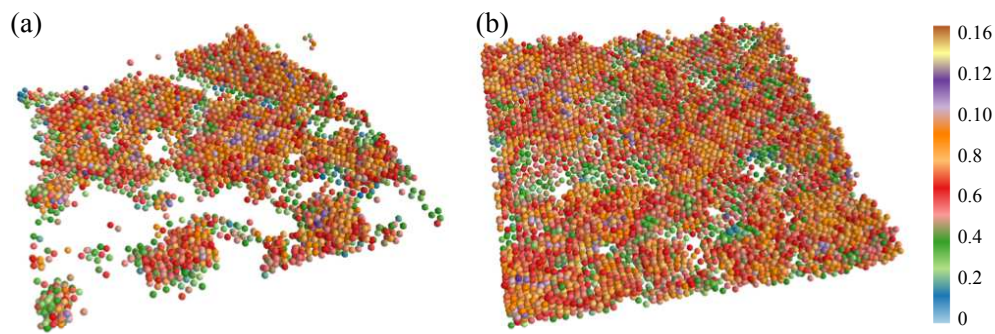}
\caption{(color online) Snapshots showing the crystallized particles on one of the two lattice substrates at density $0.6\sigma^{-3}$ (substrate not shown). The color coding is according to the $\bar{q}_{4}$ value at time $t=\tau_{D}$ for two different wall-fluid potentials: (a) the WCA potential and (b) the LJ potential with cutoff at $1.5\sigma$. }
\label{snaps}
\end{figure}

We analyze the laterally averaged (in the substrate plane) density profile $\langle\rho(z)\rangle_{x,y}$, Fig.~\ref{composition1}(b), computing the width of the first layer density peak, modeled by a Gaussian $f(z)=A\exp[-(z-\mu)^{2}/2\sigma^{2}]$ at the early stages of the growth. We notice that the WCA substrate--fluid potential induces the largest peak widths, and that the tightest peaks correspond to substrate lattices in the range of the coexistence densities. Considering that the spacing between the substrate crystalline planes in $z$ direction is $a/2$, we observe that all the early density peaks are narrow and the particles are essentially confined to a single plane. This implies that the crystallization process is in essence a 2d process at strong supersaturation, and hence it is not subject to a kinetic barrier. From this observation we conclude that heterogeneous nucleation with a non-zero contact angle in a Lennard-Jones system on a planar substrate only occurs for very specific choices of attraction range and substrate lattice spacing. 

The rapid formation of tight density peaks is accompanied by the fast growth of crystalline structures:  the first layer of high $\bar{q}_{6}$ particles is formed in about $1 \tau_{D}$ in all cases with the exception of the lowest lattice density case. Yet, the shorter the range of the attractive forces, the lower is the growth rate, in agreement with the typically longer timescales observed in the case of hard spheres. We also report that the growth rates at the initial stages of crystallization are non monotonic in the template lattice structure, suggesting that both too dense and too sparse substrates with respect to the bulk density present incommensurability barriers that slow down the layer formation.

In Fig.~\ref{composition1}(c) we show the relative fraction of solid particles in the first layer at a very early time ($t=0.25\tau_{D}$ after the wall-fluid contact) as a function of the fcc substrate density. For different wall-fluid potentials we see a significant change in the crystallinity of the first layer: hard potentials have low fractions of solid particles (below 50\%) while attractive potentials induce crystallinity in the entire first layer in a broad range of substrate lattice spacings. This is associated with a complete coverage of the substrate surface, which gradually fades out outside of the interval of coexistence densities. In the case of wall-fluid interactions with $r_{cut}=2.5\sigma$ the solid particles ratio drops rapidly only at very low lattice densities ($\rho_{\rm lattice}=0.5\sigma^{-3}$) when the lattice spacing is so large that the fluid particles penetrate through the first layer of fixed particles. 

However, even at low fractions of solid particles, we do not observe the formation of a cluster -- as it would be expected in the case of heterogeneous nucleation with a non-zero contact angle -- but rather of a network morphology (see Fig.~\ref{snaps} left image).
The lattice densities that lie in the coexistence region induce the fastest and most efficient formation of low defect number crystalline layers. Both, low and high densities lead to an increase in the crystal layer imperfections due to the mismatch between the  bulk equilibrium structure and the actual structure of fixed particles, as seen for the hard spheres case. Heterogeneous nucleation does not appear at all for the substrate lattice spacings studied here.

\begin{figure}[h!tbp]
\centering
\includegraphics[width=\columnwidth]{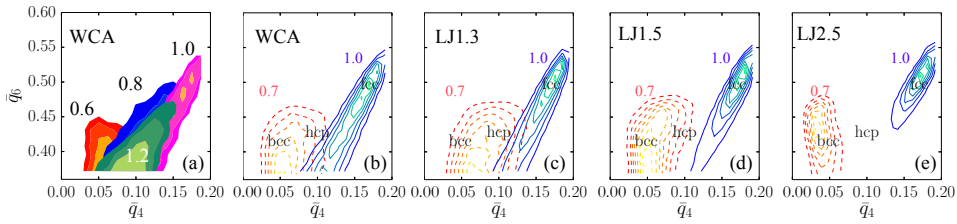}
\caption{(color online)   (a) Probability distribution functions of the average local bond order parameters $\bar{q}_{6},\bar{q}_{4}$ of the solid particles forming the first layer at time $0.25\tau_{D}$ for the WCA potential at different lattice densities (brightness indicates the position of the mode peak). Notice that the density $0.8 \sigma^{-3}$ is partly hidden by the overlaid distributions for $\rho=1.0,1.2 \sigma^{-3}$. (b-e) Probability distributions (drawn as level curves) for different wall-fluid interactions at a two different lattice densities $0.7\sigma^{-3}$ (dashed lines) and $1.0\sigma^{-3}$ (continuous lines) . The positions of the typical values of bcc, hcp and fcc stacking in the bulk LJ system \cite{Lechner:2008vz} are also indicated. }
\label{pdfs}
\end{figure}
While for long cutoff radii all the systems with lattice densities within the coexistence region show a similar growth behavior, we see that the shorter is the range of the attractive part of the potential, the more sharp is the peak in the solid fraction around some density in the coexistence region. We observe that the peak value is approximately located around the spinodal density indicated in \cite{Kuksin:2007kq} for the LJ fluid.

We then focused our analysis on the typical order that is formed at the template-fluid interface. We use the $\bar{q}_{4},\bar{q}_{6}$  probability distribution functions for the solid particles in order to represent the several possible ordered patterns that cover the template at the early stage of the crystal growth. As shown in Fig.~\ref{pdfs}(a,b), at low lattice densities (corresponding to a small number of crystalline particles) low $\bar{q}_{4}$ structures of distorted bcc type are formed; increasing the lattice density favors a more compact stacking of hcp nature. At later times the stabilization of the crystalline structure further increases the degree of local order of the initially formed layers, transforming the hcp order into fcc crystals.

The change in the range of attractive forces also contributes to distort the arrangement of the first layer of particles (Fig.\ref{pdfs}(b-e))). At low lattice densities, the potentials that are mainly repulsive show hcp ordering while the more attractive ones present bcc ordering; for higher lattice densities repulsive potentials still favor hcp-like structures while more attractive wall-fluid interactions make fcc structures more and more likely.

The different crystalline arrangements occur in different regions of thew newly formed clusters and crystal layers: as shown in Fig.~\ref{snaps} bcc particles are mainly localized on the borders of the clusters and in contact with the substrate, while hcp (and at later times, fcc) particles form the core of the crystalline clusters (Fig.~\ref{snaps}(a)) and layers (Fig.~\ref{snaps}(b)).
\newline

\section{Conclusion}

We reviewed several aspects of the thermodynamics of crystal/liquid phases and their interfaces in the hard--sphere and Lennard--Jones model
systems, as well as homogeneous and substrate--driven heterogeneous nucleation in these systems. Through the application of modern density functional methods
(especially fundamental measure functionals) a detailed theoretical understanding of the equilibrium crystal/liquid properties 
in accordance with simulation results has been achieved. We illustrated this for the particular case of the crystal(fcc,bcc)--liquid interface of 
hard spheres. For the nonequilibrium processes of homogeneous and heterogeneous nucleation a physical picture is developing mostly on the
basis of simulation results. Simple hard spheres show a precursor--mediated homogeneous nucleation scenario. Crystal growth in hard spheres at 
crystalline substrates proceeds by instantaneous growth or the classical nucleation scenario, depending on the lattice constant of the substrate.  
For the case of a Lennard--Jones system,
attractive substrate forces accelerate the formation of crystalline layers. 
The distinction between instantaneous growth and nucleation--dominated crystallization, as it has been observed for hard spheres, 
is blurred in systems with attractive interactions; the substrate is wet by a crystalline film even for highly incommensurate lattice spacings, 
making it almost impossible to study heterogeneous crystal nucleation with a non-zero contact angle in Lennard-Jones systems. 
(A similar observation has been made for Lennard Jones systems in contact with an unstructured Lennard Jones wall \cite{Gribova2011}.) 


{\bf Acknowledgments:} The authors thank the priority program SPP 1296 (``Heterogeneous Nucleation") of the German Research Foundation (DFG) for
funding through the contracts Oe 285/1-3 and Schi 853/2-1,-2 as well as the Fonds National de la Recherche Luxembourg, 
project "Crystallization INTER/DFG/11/06". \tr{The authors would like to thank M. Berghoff, 
K. Binder, A. Choudhary, D. Deb, S. Dorosz, H. Emmerich, A. H\"artel, J. Horbach, H. L\"owen, B. Nestler, H.-J. Sch\"ope, A. Tr\"oster, P. Virnau, and A. Winkler
for collaboration throughout the SPP duration.}

\begin{appendix}

\section{The excess free energy functional of FMT}
\label{app:fmt}

Fundamental measure theory (FMT) currently is the most
precise functional for the excess free energy part for the hard sphere fluid. The corresponding
excess free energy is given by
\bea
 \mathcal{F}^{\rm ex} &=& \int d^3r f^{\rm ex}(\{\mathbf{n}[\rho(\mathbf{r})]\})) \\
 \beta f^{\rm ex}(\{\mathbf{n}[\rho(\mathbf{r})]\})) &=& n_{0} \ln(1-n_{3}) + \varphi_{1}(n_{3})
\frac{n_{1} n_{2} - \mathbf{n}_{1}\cdot \mathbf{n}_{2}}{1-n_{3}} \nonumber\\
& &+ \varphi_{2}(n_{3}) \frac{3\;(-n_{2}\;\mathbf{n_{2}} \cdot \mathbf{n_{2}} + n_{2,i}n^{t}_{ij}n_{2,j}+n_{2}n^{t}_{ij}n^{t}_{ji}
- n^{t}_{ij}n^{t}_{jk}n^{t}_{ki} )}{16 \pi (1-n_{3})^{2}}\;.
\eea
Here, $f^{\rm ex}$ is the excess free energy density which is a
(local) function of a set of weighted densities $\{\mathbf{n}(\mathbf{r})\}=\{n_{0},n_{1},n_{2},n_{3}, 
\mathbf{n}_{1}, \mathbf{n}_{2}, n_{T} \}$ with four scalar, two vector and one tensorial weighted densities.
These are related to the density profile $\rho(\mathbf{r})$ by the convolutions $n_{\alpha}(\mathbf{r}) = \int d\mathbf{r}^{\prime}
\,\rho(\mathbf{r}^{\prime})\,w^{\alpha}(\mathbf{r}-\mathbf{r}^{\prime})$.
The weight functions are given by ($R=\sigma/2$ is the hard sphere radius):
\bea
w^{3}(\mathbf{r}) &=& \Theta(R - r),\nonumber\\
w^{2}(\mathbf{r}) &=& \delta(R - r),\nonumber\\
w^{1}(\mathbf{r}) &=& w^{2}(\mathbf{r})/(4\pi R),\nonumber\\
w^{0}(\mathbf{r}) &=& w^{2}(\mathbf{r})/(4\pi R^{2}),\\
\mathbf{w}^{2}(\mathbf{r}) &=& \frac{\mathbf{r}}{r}\delta(R - r),\nonumber\\
\mathbf{w}^{1}(\mathbf{r}) &=& \frac{\mathbf{w}^{2}}{4\pi R},\nonumber\\
w^{t}_{ij} &=& \frac{r_{i}r_{j}}{\mathbf{r}^{2}}\delta(R-r)\nonumber
\eea
By choosing
\bea
\varphi_{1}=1\quad \mbox{and} \quad \varphi_{2}=1
\eea
we obtain Tarazona's tensor functional~\cite{Tar00} based on the original Rosenfeld functional (RF--T) \cite{Ros89}.
The choice
\bea
\varphi_{1}&=&1 \;, \nonumber \\
\varphi_{2}&=&1-\frac{-2n_{3}+3n^{2}_{3}-2(1-n_{3})^{2}\ln (1-n_{3})}{3n^{2}_{3}}
\eea
corresponds to the tensor version of the White Bear I functional (WB--T)~\cite{Rot02,Yu02}. Finally, with
\bea
\varphi_{1}&=&1+\frac{2n_{3}-n^{2}_{3}+2(1-n_{3})\ln (1-n_{3})}{3n^{2}_{3}}\nonumber \;, \\
\varphi_{2}&=&1-\frac{2n_{3}-3n^{2}_{3}+2n^{3}_{3}+2(1-n_{3})^{2}\ln (1-n_{3})}{3n^{2}_{3}}
\eea
the tensor version of the White Bear II functional is recovered (WBII--T)~\cite{Han06}. This functional is most consistent with respect
to restrictions
imposed by morphological thermodynamics~\cite{Koen04}.

We remind the reader briefly on the construction principles of FMT: The scalar and vector densities are introduced by requiring the
correct low--density limit of the free energy for a hard sphere mixture \cite{Ros89}. The particular, analytic form of 
$f^{\rm ex}({\vect n})$ arises by imposing consistency with scaled particle arguments \cite{Ros89,Rot10} or by imposing a known bulk equation of state
\cite{Rot02,Han06}. Since in this step only bulk properties are used, there is still freedom in extending the functional to arbitrarily inhomogeneous
situations. One further demands that the functional reproduces the known free energy of a sharply peaked density distribution (0d limit) \cite{Tar97}. 
This leads to the introduction of the tensor weights as suggested in Ref.~\cite{Tar00}.

\section{Simulation details for the heterogeneous crystal growth on stretched substrates with tunable attractive potentials}
\label{simulationdetails}

We simulated $N=216\,000$ Lennard--Jones\footnote{The potential of the Lennard--Jones interaction (cut off at $r_{\rm cut}$) is defined by
$u(r)=u_{\rm LJ}(r)-u_{\rm LJ}(r_{\rm cut})$ for $r<r_{\rm cut}$ and 0 otherwise, with $u_{\rm LJ}(r)=4\epsilon[(\sigma/r)^{12}-(\sigma/r)^6 ]$.} fluid particles at temperature $k_{B}T=0.8\epsilon$ and density $\rho=0.95\sigma^{-3}$ confined by two fcc walls of fixed particles of surface $A=30\times30 a^{2}$ where $a$ is the fcc lattice spacing $a=\sqrt[3]{4/\rho_{\rm lattice}}$. The substrates expose the $(100)$ orientation to the fluid, in contrast with the $(111)$ studied for hard-spheres case in \cite{Dorosz:2012ka}. The wall particles form three crystalline layers per surface and interact with the fluid particles with different possible potentials, all based on the pair Lennard--Jones interaction, as illustrated in Fig.~\ref{composition1}(a): we consider a case for which the wall-liquid interaction is the same as the liquid-liquid interaction (purely cut and shifted Lennard-Jones interaction with cutoff $r_{\rm cut}=2.5\sigma$); then, we pick the limit case for which no attractive force is present between the substrate and the liquid (cutting and shifting the LJ potential at $r_{\rm min}=2^{1/6}\sigma$) corresponding to the Weeks-Chandler-Andersen (WCA) potential and finally we choose intermediate truncated and shifted potentials where we limit the contribution of the attractive part choosing short cutoff radii $r_{\rm cut}=1.3,1.5\sigma$.
We perform isochoric Langevin dynamics simulations with a time step $\Delta t=0.01 \sqrt{m\sigma^{2}/\epsilon}$ and friction coefficient $\gamma=0.01 \Delta t^{-1}$ and track the crystallization process using the averaged local bond order parameters $\bar{q}_{4},\bar{q}_{6}$ proposed in \cite{Lechner:2008vz}. Their definition requires the computation of the complex vector $q_{l}(i)$
\begin{equation}
q_{lm}(i)=\frac{1}{N_{b}(i)}\sum_{j=1}^{N_{b}(i)}Y_{lm}(\bm{r}_{ij}) \;, 
\end{equation}
where $N_{b}(i)$ corresponds to the number of nearest neighbors of particle $i$ and $Y_{lm}(\bm{r}_{ij})$ reads as the spherical harmonics. Averaging over the neighbors of particle $i$ and particle $i$ itself
\begin{equation}
\bar{q}_{lm}(i)=\frac{1}{\tilde{N}_{b}(i)}\sum_{k=0}^{\tilde{N}_{b}(i)}q_{lm}(k),
\end{equation}
and summing over all the harmonics we finally get
\begin{equation}
\bar{q}_{l}(i)=\sqrt{\frac{4\pi}{2l+1}\sum_{m=-l}^{l}|\bar{q}_{lm}(i)|^{2}} \;.
\end{equation}

\end{appendix}

\end{document}